
%
\input phyzzx
\def\today{
 \ifcase\month \or January \or February \or March \or April
               \or May \or June \or July \or August
               \or September \or October \or November \or December \fi
 \space \number\day, \number\year}
\def\todaywithoutYear{
 \ifcase\month \or Jan. \or Feb. \or Mar. \or Apr.
               \or May  \or Jun. \or Jul. \or Aug.
               \or Sep. \or Oct. \or Nov. \or Dec. \fi
 \number\day}
\Pubnum={ INS-Rep.-1038 }
\date={   June 1994 }
\def\eqname#1{\relax \ifnum\equanumber<0
     \xdef#1{{\rm(\number-\equanumber)}}\global\advance\equanumber by -1
    \else \global\advance\equanumber by 1
      \xdef#1{{\rm(\number\equanumber)}} \fi}
\def\NextPage{\vskip 36pt}

%
%
\titlepage
\vskip 42pt
\title{ Non-critical String Field Theory  \nextline
        with Non-orientable String Interactions
      }
\vskip 16pt
\author{ Yoshiyuki WATABIKI }
\vskip 16pt
\address{ Institute for Nuclear Study, University of Tokyo,
          \nextline
          Tanashi, Tokyo 188, Japan }
\vskip 36pt
%
%

\font\BIGr=cmr10 scaled \magstep2

\def\sp(#1){ \noalign{\vskip #1pt} }

\def\der{\partial}

\def\e{\varepsilon}
\def\derover(#1,#2){ { \der_{#1} \over \der_{#2} } }

\def\dim#1{ { \hbox{dim} [ #1 ] } }
\def\bra#1{ \langle #1 | }
\def\ket#1{ | #1 \rangle }
\def\vac{ \bra{{\rm vac}} }
\def\cuum{ \ket{{\rm vac}} }

\def\k{\kappa}
\def\inv#1{ {1 \over #1} }
\def\Real{ {\Re e} }

\def\define{ \mathrel{\mathop\equiv^{\rm def}} }
\def\intdz#1{ { \oint\limits_{|z|=#1} \! \! {d z \over 2 \pi i z} } }
\def\intdzeta{ { \int\limits_{- i \infty}^{+ i \infty}
               {d \zeta \over 2 \pi i} } }
\def\void{\hskip 10pt}
\def\cc{t} 
\def\t{d}
\def\T{D}
\def\gx{g'}   
\def\gxx{g''} 
\def\Gx{G'}   
\def\Gxx{G''} 
\def\H{ {\cal H} }
\def\Htot{ {\hat \H}(\k) } 
\def\HTOT{ \H(\cc) } 

%

%
%
\abstract{
We study
the non-critical string field theory with non-orientable string interactions
by using the transfer matrix formalism in the dynamical triangulation.
For any value of $c$ (total central charge of matter),
we have constructed the non-orientable string field theory
at the discrete level.
Two coupling constants $G$ and $\Gx$
which satisfy the relation $(\Gx)^3 = G \Gx$
are introduced,
where $G$ counts the number of untwisted handles of the world-sheet
while $\Gx$ counts the number of cross-caps.
In the case of the non-critical string theory
which correspond to multicritical one matrix models
(including $c=0$ case),
we have succeeded in taking the continuum limit,
and then have obtained the continuous string field theory.
}
%
%
\endpage             
%
%
%
\REF\SFT{    M. Kaku and K. Kikkawa,
                Phys.\ Rev.\ {\bf D10} (1974) 1110, 1823;
             W. Siegel,
                Phys.\ Lett.\ {\bf B151} (1985) 391, 396;
             H. Hata, K. Itoh, T. Kugo, H. Kunitomo and K. Ogawa,
                Phys.\ Lett.\ {\bf B172} (1986) 186, 195;
             A. Neveu and P. West,
                Phys.\ Lett.\ {\bf B168} (1986) 192;
             E. Witten,
                Nucl.\ Phys.\ {\bf B268} (1986) 253.
             }
\REF\DT{     V. A. Kazakov,
                Phys.\ Lett.\ {\bf B150} (1985) 282;
             F. David,
                Nucl.\ Phys.\ {\bf B257} (1985) 45, 543;
             J. Ambj\o rn, B. Durhuus and J. Fr\"ohlich,
                Nucl.\ Phys.\ {\bf B257} (1985) 433;
             V. A. Kazakov, I. K. Kostov and A. A. Migdal,
                Phys.\ Lett.\ {\bf B157} (1985) 295;
             D. Boulatov, V. A. Kazakov, I. K. Kostov and A. A. Migdal,
                Phys.\ Lett.\ {\bf B174} (1986) 87;
                Nucl.\ Phys.\ {\bf B275} (1986) 641.
             }
\REF\KKMW{   H. Kawai, N. Kawamoto, T. Mogami and Y. Watabiki,
                Phys.\ Lett.\ {\bf B306} (1993) 19.
             }
\REF\IK{     N. Ishibashi and H. Kawai,
                Phys.\ Lett.\ {\bf B314} (1993) 190;
                              {\bf B322} (1994) 67.
             }
\REF\Klebanov{  S.S. Gubser and I.R. Klebanov,
                Nucl.\ Phys.\ {\bf B416} (1994) 827.
             }
\REF\Biki{   Y. Watabiki,
                INS-Rep.-1017 December (1993).
             }
\REF\Kazakov{   V.A. Kazakov,
                Mod.\ Phys.\ Lett.\ {\bf A4} (1989) 2125.
             }
\def\Refmark(#1){$\,$[#1]}
\def\CMPrefmark(#1){ [#1]}

%

%
%
%
\def\FigPeel{1}
\def\FigPeelSkin{2}
 \def\FigPeelSkinI{2a}
 
 \def\FigPeelSkinIII{2c}
 \def\FigPeelSkinIV{2d}
 \def\FigPeelSkinV{2e}
\def\FigPeelTri{3}

%
\topskip 30pt
%
%
\par
The (critical) string theories are hopeful candidates for unified theory.
In order to predict the phenomenology at low energy,
we have to know the non-perturbative effects of the string interactions.
With the aim of understanding the non-perturbative effects,
the string field theories were constructed.\NPrefmark{\SFT}
However, no one could extract the information of the non-perturbative effects
because the string field theories are too complicated for us to analyze.
Recently,
in the framework of the dynamical triangulation,\NPrefmark{\DT}
non-critical string field theories have been investigated
in refs.\ [{\KKMW}, {\IK}, {\Klebanov}, {\Biki}].
As was shown in ref.\ [{\KKMW}],
the transfer matrix formalism for $c=0$ non-critical string theory
is powerful to analyze the fractal structure of quantized world-sheet.
In ref.\ [{\IK}] the authors pushed the transfer matrix formalism forward
and proposed the non-critical string field theories.
Since the bosonic (critical) string theory is equivalent to
the $c=25$ non-critical string theory,
it is interesting to investigate the non-critical string field theories
as toy models of the string field theories.
\par
The purpose of the present paper is to construct
the non-critical string field theory
which has non-orientable string interactions.
We here use the systematic construction proposed by ref.\ [{\Biki}].
In other words, it is shown in this paper that
the construction of the non-critical string field theory
given by ref.\ [{\Biki}] is also valid
for the string field theories with non-orientable string interactions.
We have succeeded in constructing
the non-critical non-orientable string field theory at the discrete level
for any value of $c$ (including $c > 1$ case),
by using the transfer matrix formalism
in the dynamical triangulation of non-orientable surfaces.
Moreover, we have also succeeded in taking the continuum limit
for the models which correspond to
$m$-th multicritical one matrix models\NPrefmark{\Kazakov}
(including $c=0$ case).
\par
In ref.\ [{\Biki}] we constructed the string field theory
by decomposing a triangulated surface like peeling an apple.
The \lq peeling decomposition'
removes each triangle on the boundary sequentially
like Fig.\ {\FigPeel}
and identifies each removal by $(1/l)$-step decomposition,
where $l$ is the number of links on the boundary,
i.e., the length of the boundary.
The \lq step' plays the role of proper time in the field theory.
Note that in the peeling decomposition no string is created from the vacuum.
A finite-step peeling decomposition is obtained by performing
the minimal-step one over and over again.
As is illustrated in Figs.\ {\FigPeelSkin},
there are five kinds of minimal-step peeling decompositions
when the world-sheet is orientable or non-orientable.
We consider each decomposition in Figs.\ {\FigPeelSkin}
as $(1/l)$-step decomposition
because the initial length of the boundary is $l$.
Note that
the deformations in
Figs.\ {\FigPeelSkinIII}, {\FigPeelSkinIV}, and {\FigPeelSkinV}
make an untwisted handle, a cross-cap, and a twisted handle,
respectively.
%
%
\par
In the beginning
we consider the dynamical triangulation of non-orientable surfaces
with no matter fields.
Let ${\hat \Psi}^\dagger (l)$ and ${\hat \Psi} (l)$ be operators,
which creates and annihilates one closed string with length
$l$ ($= 1,2,3,\ldots$),
respectively.
Each string has one marked link
in order to indicate the present peeling point and the peeling direction.
Their commutation relations are assumed to be
$$ \eqalign{ \sp(2.0)
[ \, {\hat \Psi} (l) \, , \, {\hat \Psi}^\dagger (l') \, ]  \ = \
\delta_{l,l'} \, ,
\quad
\hbox{otherwise} \ = \ 0 \, .
\cr\sp(3.0)} \eqn\EqCommuteDisI $$
The vacuum states, $\cuum$ and $\vac$, satisfy
${\hat \Psi} (l) \cuum = \vac {\hat \Psi}^\dagger (l) = 0$ (for any $l \ge 1$)
and
$\langle {\rm vac} | {\rm vac} \rangle  = 1$.
\par
Now, let us consider the Hamiltonian, $\Htot$,
where $\k$ counts the number of triangles by putting it on each triangle.
The $(1/l)$-step deformed wave function,
${\hat \Psi}^\dagger(l) + \delta_{1/l} {\hat \Psi}^\dagger(l)$,
is derived from the Hamiltonian $\Htot$ by
$$ \eqalign{ \sp(2.0)
\delta_{1/l} \, {\hat \Psi}^\dagger (l)
\ = \
- \, [ \, {1 \over l} \, \Htot \, , \, {\hat \Psi}^\dagger (l) \, ] \, .
\cr\sp(3.0)} \eqn\EqOneOverLHamilton $$
Inversely speaking,
the form of $\Htot$ is determined from $\delta_{1/l} {\hat \Psi}^\dagger(l)$
through eq.\ {\EqOneOverLHamilton}.
Since no string is created from the vacuum in the peeling decomposition,
the vacuum state is invariant under the time evolution,
i.e.,
$$ \eqalign{ \sp(2.0)
\Htot \, \cuum  \ = \  0  \, .
\cr\sp(3.0)} \eqn\EqHvac $$
\par
In the dynamical triangulation
all possible triangulated surfaces are summed up.
Since the decomposition of the surface is equivalent to
the deformation of the wave function,
we sum up all peeling decompositions in Figs.\ {\FigPeelSkin},
and then we find
$$ \eqalign{ \sp(2.0)
{\hat \Psi}^\dagger (l) \, + \, \delta_{1/l} \, {\hat \Psi}^\dagger (l)
\  =  \
&  \k {\hat \Psi}^\dagger (l+1)
\, + \, ( 1 - \delta_{l,1} ) \,
   \sum_{l'=0}^{l-2} {\hat \Psi}^\dagger (l') {\hat \Psi}^\dagger (l-l'-2)
\cr\sp(4.0)
&  + \, 2 g \, \sum_{l'=1}^\infty {\hat \Psi}^\dagger (l+l'-2)
     l' {\hat \Psi} (l')
\cr\sp(4.0)
&  + \, \gx \, (l-1) \, {\hat \Psi}^\dagger (l-2)
\, + \, 2 \gxx \, \sum_{l'=1}^\infty {\hat \Psi}^\dagger (l+l'-2)
     l' {\hat \Psi} (l') \, ,
\cr\sp(3.0)} \eqn\EqDeform $$
where ${\hat \Psi}^\dagger (l=0) = 1$ is introduced
in order to simplify the form of the right-hand side.
Each term in the right-hand side of {\EqDeform}
corresponds to the deformation by
Figs.\ {{\FigPeelSkinI} $\sim$ {\FigPeelSkinV}}, respectively.
The factors $l'$, $l-1$ and $l'$
in the third, fourth and fifth terms
in the right-hand side of {\EqDeform} are necessary
because there are $l'$, $l-1$, and $l'$ types of figures for
Figs.\ {{\FigPeelSkinIII}, {\FigPeelSkinIV} and {\FigPeelSkinV},
respectively.
The coupling constants $g$, $\gx$ and $\gxx$ are introduced
in order to count the number of untwisted handles, that of cross-caps,
and that of twisted handles, respectively.
{}From {\EqOneOverLHamilton}, {\EqHvac} and {\EqDeform},
the form of the Hamiltonian is uniquely determined as
$$ \eqalign{ \sp(2.0)
\Htot \ = \
&\sum_{l=1}^\infty \bigl\{ \,
 {\hat \Psi}^\dagger (l) \, - \, \k {\hat \Psi}^\dagger (l+1)
 \, \bigr\} \, l \, {\hat \Psi} (l)
\cr\sp(4.0)
&- \, \sum_{l=2}^\infty \, \sum_{l'=0}^{l-2} \bigl\{ \,
      {\hat \Psi}^\dagger (l') {\hat \Psi}^\dagger (l-l'-2)
      \, + \, \gx \, {\hat \Psi}^\dagger (l-2)
   \, \bigr\} \, l \, {\hat \Psi} (l)
\cr\sp(4.0)
&- \, ( g + \gxx ) \, \sum_{l=1}^\infty \sum_{l'=1}^\infty
 {\hat \Psi}^\dagger (l+l'-2) \, l \, {\hat \Psi} (l) \, l' \, {\hat \Psi} (l')
\, .
\cr\sp(3.0)} \eqn\EqHDisI $$
Here, one should note the following well-known mathematical properties:
\item{  i) }{
one twisted handle is equivalent to two cross-caps,
}
\item{ ii) }{
one untwisted handle is equivalent to one twisted handle
under the existence of a cross-cap.
}
\par{\noindent}Then, all the two-dimensional manifolds
are classified without duplication as follows:
orientable surfaces
with $N$ ($\ge 0$) boundaries and $h$ ($\ge 0$) handles,
and
non-orientable surfaces
with $N$ ($\ge 0$) boundaries and $k$ ($\ge 1$) cross-caps.
Therefore,
the topological properties {\it i}) and {\it ii}) require
$$ \eqalign{ \sp(2.0)
  \gxx  \ = \  (\gx)^2 \, ,
\qquad
  g \, \gx  \ = \  \gxx \, \gx \, ,
\cr\sp(3.0)} \eqn\EqCouplingDis $$
respectively.
\par
Expanding the amplitudes in terms of $g$ and $\gx$,
we obtain each contribution from orientable and non-orientable surfaces.
Thus, the transfer matrices,
${\hat T}_{M,N}^{(h)}$ with $h$ handles
and
${\hat T}_{M,N}^{\prime (k/2)}$ with $k$ cross-caps,
are obtained by
$$ \eqalign{ \sp(2.0)
& { \delta^{M+N} \over
    \delta {\hat J}^\dagger (l'_1) \, \cdots \, \delta {\hat J}^\dagger (l'_M)
    \delta {\hat J} (l_1) \, \cdots \, \delta {\hat J} (l_N) } \,
  \ln {\hat Z} [{\hat J}^\dagger,{\hat J};\k;\t]
  \Big|_{{\hat J}^\dagger={\hat J}=0}
\cr\sp(6.0)
& = \
  ( g + \gxx )^{N-1} \, \bigl\{ \,
  \sum_{h=0}^\infty \, g^h \,
  {\hat T}_{M,N}^{(h)} (l'_1,\ldots,l'_M;l_1,\ldots,l_N;\k;\t)
\cr\sp(4.0)
&\phantom{ =  \  ( g + \gxx )^{N-1} \, \bigl\{ \, }
  + \, \sum_{k=1}^\infty \, \gx^k \,
  {\hat T}_{M,N}^{\prime (k/2)} (l'_1,\ldots,l'_M;l_1,\ldots,l_N;\k;\t)
 \, \bigr\}
\cr\sp(3.0)} \eqn\EqGenTMDisI $$
with
$$ \eqalign{ \sp(2.0)
   {\hat Z} [{\hat J}^\dagger,{\hat J};\k;\t]  \ \define \
   \vac \,
   e^{\sum_{l'} \! {\hat J}^\dagger (l') {\hat \Psi} (l')} \,
   e^{- \t \Htot} \,
   e^{\sum_{l} \! {\hat \Psi}^\dagger (l) {\hat J} (l)} \,
   \cuum  \, ,
\cr\sp(3.0)} \eqn\EqGenFunDisI $$
where
$N$ ($\ge 1$) and $M$ ($\ge 0$) are the number of initial string boundaries
and that of final string boundaries, respectively.
Note that the transfer matrices in the peeling decomposition
have at least one initial string boundary ($N \ge 1$).
Thus, we have completed the construction of
the discretized $c=0$ non-orientable string field theory.
\par
Before taking the continuum limit,
we introduce the discrete Laplace transformation
in order to avoid ambiguity when we take the continuum limit.
The discrete Laplace transformation and its continuum limit
are explained in detail in the appendix A of ref.\ [{\Biki}], for example.
We define the discrete Laplace transformation of the wave functions by
$$ \eqalign{ \sp(2.0)
  {\hat \Psi}^\dagger (x)  \ \define \
  {\hat \Psi}^\dagger(l=0) \, + \,
  \sum_{l=1}^\infty \, x^l \, {\hat \Psi}^\dagger (l) \, ,
  \qquad
  {\hat \Psi} (y)  \ \define \
  \sum_{l=1}^\infty \, y^l \, {\hat \Psi} (l) \, .
\cr\sp(3.0)} \eqn\EqLaplacePsiDis $$
We here assume that
the wave functions ${\hat \Psi}^\dagger(x)$ and ${\hat \Psi}(y)$ are analytic
in the region $|x| \le x_c$ and $|y| \le y_c$,
where $x_c$ and $y_c$ are the convergence radii of
${\hat \Psi}^\dagger(x)$ and ${\hat \Psi}(y)$, respectively.
According to ref.\ [{\Biki}],
we redefine the wave function as
$$ \eqalign{ \sp(2.0)
{\hat \Phi}^\dagger (x,\k)
\  \define \
{\hat \Psi}^\dagger (x) \, - \, {\hat \lambda} ( x , \k ) \, ,
\qquad
{\hat \lambda} ( x , \k )  \ = \
{1 \over 2 x^2} \bigl( 1 - {\k \over x} \bigr) \, .
\cr\sp(3.0)} \eqn\EqDefPhiDagger $$
If and only if $x_c y_c = 1$,
the Hamiltonian {\EqHDisI} is expressed by
${\hat \Phi}^\dagger(x,\k)$ and ${\hat \Psi}(y)$ as
$$ \eqalign{ \sp(2.0)
  \Htot \ = \  - \intdz{z_c} \, \bigl\{ \,
& {\hat \omega} (z,\k) \, {\hat \Psi} (\inv{z})
  \, - \, z^2 ( {\hat \Phi}^\dagger (z,\k) )^2
       \, z {\der \over \der z} \, {\hat \Psi} (\inv{z})
\cr\sp(0.1)
& + \, ( g + \gxx ) \, z^2 {\hat \Phi}^\dagger (z,\k)
                    \, ( z {\der \over \der z} \, {\hat \Psi} (\inv{z}) )^2
\cr\sp(6.0)
& + \, \gx  \, z^2 {\hat \Phi}^\dagger (z,\k)
            \, {\der \over \der z} \, ( z^2 {\der \over \der z}
            \, {\hat \Psi} (\inv{z}) ) \, \bigr\} \, ,
\cr\sp(3.0)} \eqn\EqHDisII $$
where
$$ \eqalign{ \sp(2.0)
{\hat \omega} ( x , \k ) \ = \
- \, x {\der \over \der x} \bigl\{ \,
  {1 \over 4 x^2} \bigl( 1 - {\k \over x} \bigr)^2
  \, + \, ( 1 - \gx ) \, {\k \over x}
\, \bigr\} \, .
\cr\sp(3.0)} \eqn\EqOmegaDis $$
%
%
\par
Now let us consider the continuum limit,
which is taken by $\e \rightarrow 0$ with
$$ \eqalign{ \sp(2.0)
x  \  = \ x_c \, e^{- \e \xi} \, ,
\qquad
y  \  = \ y_c \, e^{- \e \eta} \, ,
\qquad
\k \  = \ \k_c \, e^{- \e^2 c_0 \cc} \, ,
\cr\sp(3.0)} \eqn\EqConLimitxyk $$
where $\cc$ is the cosmological constant at the continuous level.
The continuum limit of the wave functions is assumed to be
$$ \eqalign{ \sp(2.0)
&\Psi^\dagger (\xi)  \ = \  \lim_{\e \rightarrow 0}
 c_1 \, \e^\dim{\Psi^\dagger} \, {\hat \Psi}^\dagger (x) \, ,
\qquad
 \Psi (\eta)        \ = \   \lim_{\e \rightarrow 0}
 c_2 \, \e^\dim{\Psi} \, {\hat \Psi} (y)  \, ,
\cr\sp(4.0)
&\Phi^\dagger (\xi,\cc)  \ = \  \lim_{\e \rightarrow 0}
 c_3 \, \e^\dim{\Phi^\dagger} \, {\hat \Phi}^\dagger (x,\k) \, .
\cr\sp(3.0)} \eqn\EqConLimitWave $$
We also assume that
$$ \eqalign{ \sp(2.0)
  \HTOT  \ = \
  \lim_{\e \rightarrow 0} c_4 \, \e^\dim{\H} \, \Htot  \, ,
\qquad
  \T  \ = \
  \lim_{\e \rightarrow 0} c_5 \, \e^\dim{\T} \, \t \, ,
\cr\sp(3.0)} \eqn\EqConLimitDH $$
and
$$ \eqalign{ \sp(2.0)
 G    \ = \  \lim_{\e \rightarrow 0} c_6 \, \e^{\dim{G}} \, g  \, ,
\quad
 \Gx  \ = \  \lim_{\e \rightarrow 0} c_7 \, \e^{\dim{\Gx}} \, \gx  \, ,
\quad
 \Gxx \ = \  \lim_{\e \rightarrow 0} c_8 \, \e^{\dim{\Gxx}} \, \gxx  \, .
\cr\sp(3.0)} \eqn\EqConLimitG $$
Here, $\dim{P}$ is the dimension of $P$ at the continuous level
in the unit of $\dim{\e} = 1$.
The coefficients $c_0$ $\sim$ $c_8$ are arbitrary positive numbers.
$\Psi^\dagger(\xi)$ and $\Psi(\eta)$ 
are analytic in the region
$0 \le \Real(\xi)$ and $0 \le \Real(\eta)$, 
because
${\hat \Psi}^\dagger(x)$ and ${\hat \Psi}(y)$ 
are analytic in the region
$|x| \le x_c$ and $|y| \le y_c$. 
In order to take the continuum limit,
we let $\e \rightarrow 0$ and $\t \rightarrow \infty$ while $\T$ is fixed.
Therefore, the continuum limit can be taken if and only if $\dim{\T} > 0$.
As was shown in ref.\ [{\Biki}],
the condition $\dim{\T} > 0$ is satisfied if and only if
$x_c  = (3^{1/4} - 3^{-1/4}) / 2$ and $\k_c = 3^{1/4} / 6$
in the case of the decomposition given by Figs.\ {\FigPeelSkin}.
Then, we find
$$ \eqalign{ \sp(2.0)
& \dim{\Phi^\dagger}  \ = \  \dim{\Psi^\dagger}  \ = \  - \, {3 \over 2} ,
  \qquad
  \dim{\Psi}  \ = \  {5 \over 2} ,
\cr\sp(6.0)
& \dim{\T}  \ = \  - \, \dim{\H}  \ = \  {1 \over 2} ,
  \qquad
  \dim{G}  \ = \  2 \, \dim{\Gx}  \ = \  \dim{\Gxx}  \ = \  - \, 5 .
\cr\sp(3.0)} \eqn\EqDimension $$
In order to simplify the form of the equations,
we have set the coefficients as
$c_0  = ( 3+\sqrt{3} )^2 / 16 $,
$c_1 = 1/c_2 = c_3  = 2 / (1+\sqrt{3})^{5/2}$,
$c_4  = 1/c_5 = 2 \sqrt{3} / (1+\sqrt{3})^{1/2}$,
$c_6 = c_7^2 = c_8 = x_c^2 c_3 c_4$.
As a result, we obtain
the continuum limit of {\EqDefPhiDagger} as
$$ \eqalign{ \sp(2.0)
  \Phi^\dagger (\xi)
  \ = \
  \Psi^\dagger (\xi) \, - \, \lambda(\xi) \, ,
\qquad
  \lambda (\xi)
  \ = \
  \lim_{\e \rightarrow 0} c_1 \, \e^\dim{\Psi^\dagger} \,
  {\hat \lambda} ( x , \k )
  \, ,
\cr\sp(3.0)} \eqn\EqDefPhiDaggerCon $$
and that of the commutation relations {\EqCommuteDisI} as
$$ \eqalign{ \sp(2.0)
& [ \, \Psi (\eta) \, , \, \Psi^\dagger (\xi) \, ]  \ = \
  [ \, \Psi (\eta) \, , \, \Phi^\dagger (\xi) \, ]  \ = \
  \delta ( \eta , \xi ) \, ,
\quad
\hbox{otherwise} \ = \ 0 \, ,
\cr\sp(3.0)} \eqn\EqCommuteCon $$
where $\delta(\eta,\xi) = 1/(\eta+\xi)$
is the continuous Laplace transformation of $\delta(L-L')$.
{}From {\EqHDisII} and {\EqOmegaDis},
we can uniquely determine the Hamiltonian at the continuous level as
$$ \eqalign{ \sp(2.0)
  \HTOT \ = \  - \, \intdzeta \, \bigl\{ \,
& \omega (\zeta,\cc) \, \Psi (- \zeta)
  \, + \, ( \Phi^\dagger (\zeta) )^2
       \, {\der \over \der \zeta} \, \Psi (- \zeta)
\cr\sp(0.1)
& + \, ( G + \Gxx ) \, \Phi^\dagger (\zeta)
                    \, ( {\der \over \der \zeta} \, \Psi (- \zeta) )^2 \,
  + \, \Gx  \, \Phi^\dagger (\zeta)
            \, {\der^2 \over \der \zeta^2} \, \Psi (- \zeta) \, \bigr\} \, ,
\cr\sp(3.0)} \eqn\EqHCon $$
where
$$ \eqalign{ \sp(2.0)
  \omega (\xi,\cc)  \ = \  3 \xi^2 \, - \, {3 \over 4} \cc \, .
\cr\sp(3.0)} \eqn\EqOmegaCon $$
When $\Gx=0$, the Hamiltonian {\EqHCon} is consistent with
the result of ref.\ [{\IK}].
We also find that the condition {\EqCouplingDis} becomes
$$ \eqalign{ \sp(2.0)
  \Gxx  \ = \  (\Gx)^2 \, ,
\qquad
  G \, \Gx  \ = \  \Gxx \, \Gx \, .
\cr\sp(3.0)} \eqn\EqCouplingCon $$
The continuum limit of {\EqGenTMDisI} is
$$ \eqalign{ \sp(2.0)
& { \delta^{M+N} \over
    \delta J^\dagger (\eta_1) \, \cdots \, \delta J^\dagger (\eta_M)
    \delta J (\xi_1) \, \cdots \, \delta J (\xi_N) } \,
  \ln Z^{\rm univ} [J^\dagger,J;\cc;\T] \,
  \Big|_{J^\dagger=J=0}
\, + \, \delta_{M,0} \, \delta_{N,1} \, \lambda (\xi_1)
\cr\sp(6.0)
& = \
  ( G + \Gxx )^{N-1} \, \bigl\{ \,
  \sum_{h=0}^\infty \, G^h \,
  T_{M,N}^{(h)} (\eta_1,\ldots,\eta_M;\xi_1,\ldots,\xi_N;\cc;\T)
\cr\sp(4.0)
&\phantom{ =  \  ( G + \Gxx )^{N-1} \, \bigl\{ \, }
  + \, \sum_{k=1}^\infty \, \Gx^k \,
  T_{M,N}^{\prime (k/2)} (\eta_1,\ldots,\eta_M;\xi_1,\ldots,\xi_N;\cc;\T)
 \, \bigr\}
\cr\sp(3.0)} \eqn\EqGenTMCon $$
with
$$ \eqalign{ \sp(2.0)
  Z^{\rm univ} [J^\dagger,J;\cc;\T]  \
  \define \
  \vac \,
  e^{\int \! ( d\zeta / 2 \pi i ) J^\dagger(\zeta) \Psi(-\zeta)} \,
  e^{- \T \HTOT} \,
  e^{\int \! ( d\zeta / 2 \pi i ) \Phi^\dagger(\zeta) J(-\zeta)} \,
  \cuum  \, .
\cr\sp(3.0)} \eqn\EqGenFunCon $$
The vacuum state satisfies
$\Psi (\eta) \cuum = \vac \Psi^\dagger (\xi) = 0$
(for any $\Real(\eta) \ge 0$ and any $\Real(\xi) \ge 0$).
Since $\dim{\Phi^\dagger} = \dim{\Psi^\dagger} = - 3/2$,
the $\lambda(\xi)$ defined in {\EqDefPhiDaggerCon}
is divergent but is independent of the cosmological constant $\cc$.
Therefore, we have used the notations,
$\lambda(\xi)$ and $\Phi^\dagger(\xi)$
instead of
$\lambda(\xi,\cc)$ and $\Phi^\dagger(\xi,\cc)$.
\par
Next let us consider the universality
by altering the definition of the dynamical triangulation slightly.
First,
we consider to replace regular triangles on triangulated surfaces
with regular $n$-polygons.
The discretized Hamiltonian has the same form as
{\EqHDisII} with {\EqOmegaDis}
except for the replacement of $\k$ by $\k_n / z^{n-3}$.
Especially, $\k = \k_3$.
After taking the continuum limit,
the same Hamiltonian {\EqHCon} is obtained.
Second,
we consider not to introduce
the two-folded parts like in Figs.\ {\FigPeelTri}.
Then, we obtain the discretized Hamiltonian as
$$ \eqalign{ \sp(2.0)
\Htot \ = \
&\sum_{l=1}^\infty \bigl\{ \,
 {\hat \Psi}^\dagger (l) \, - \, \k {\hat \Psi}^\dagger (l+1)
                  \, - \, 2 \k (1-\delta_{l,1}) \, {\hat \Psi}^\dagger (l-1)
\cr\sp(4.0)
&\phantom{\sum_{l=1}^\infty \bigl\{ \, }
 \, - \, \k \, \sum_{l'=1}^{l}
         \bigl( {\hat \Psi}^\dagger (l') {\hat \Psi}^\dagger (l-l'+1)
                \, + \, \gx \, {\hat \Psi}^\dagger (l+1) \bigr)
\cr\sp(4.0)
&\phantom{\sum_{l=1}^\infty \bigl\{ \, }
 \, - \, \k \, \delta_{l,3} \, - \, \k \, ( 1 + \gx ) \, \delta_{l,1}
 \, \bigr\} \, l \, {\hat \Psi} (l)
\cr\sp(4.0)
&- \, ( g + \gxx ) \, \k \, \sum_{l=1}^\infty \sum_{l'=1}^\infty
 {\hat \Psi}^\dagger (l+l'+1) \, l \, {\hat \Psi} (l) \, l' \, {\hat \Psi} (l')
\, .
\cr\sp(3.0)} \eqn\EqHNoSkinDis $$
We obtain the same Hamiltonian {\EqHCon}
after taking the continuum limit of {\EqHNoSkinDis}.
Therefore, the form of the Hamiltonian at the continuous level, {\EqHCon},
is universal.
\par
The amplitudes with $N \ge 1$ are obtained by
$$ \eqalign{ \sp(2.0)
& F_N^{(h)} ( \xi_1, \ldots, \xi_N ; \cc )
  \ = \
  \lim_{\T \rightarrow \infty}
  T_{M=0,N}^{(h)} ( \void ; \xi_1, \ldots, \xi_N ; \cc ; \T ) \, ,
\cr\sp(6.0)
& F_{N}^{\prime (k/2)} ( \xi_1, \ldots, \xi_N ; \cc )
  \ = \
  \lim_{\T \rightarrow \infty}
  T_{M=0,N}^{\prime (k/2)} ( \void ; \xi_1, \ldots, \xi_N ; \cc ; \T ) \, ,
\cr\sp(3.0)} \eqn\EqGenAmpCon $$
where the blanks between \lq$($' and \lq$;$'
mean that there are no final string states.
The explicit forms of the amplitudes are calculated
by using {\EqDefPhiDaggerCon} $\sim$ {\EqGenFunCon} and {\EqGenAmpCon},
and are obtained as, for example,
$$ \eqalign{ \sp(2.0)
& F_{1}^{\prime (1/2)} ( \xi ; \cc )
  \ = \
  { 3 \over 4 } \,
  { \xi + {1 \over 2} \, \cc^{1/2}
    - \sqrt{2 \over 3} \, \cc^{1/4} \sqrt{\xi + \cc^{1/2}}
    \over
    ( \xi - {1 \over 2} \, \cc^{1/2} ) \, ( \xi + \cc^{1/2} ) }  \, ,
\cr\sp(6.0)
& F_{1}^{\prime (1)} ( \xi ; \cc )
  \ = \
  { 1 \over 6 \cc } \,
  ( \xi - {1 \over 2} \, \cc^{1/2} )^{-3} \, ( \xi + \cc^{1/2} )^{-5/2} \,
  \bigl\{ \, \xi^4 + \xi^3 \,\cc^{1/2} - {39 \over 8} \, \xi^2 \, \cc
\cr\sp(4.0)
& \phantom{ F_{1}^{\prime (1)} ( \xi ; \cc )  \ = \  }
  - {43 \over 8} \, \xi \, \cc^{3/2} - {97 \over 32} \, \cc^2
  + {3^{5/2} \over 2^{3/2}} \, \cc^{5/4} \,
    ( \xi + {1 \over 2} \, \cc^{1/2} ) \sqrt{\xi + \cc^{1/2}} \, \bigr\}  \, .
\cr\sp(3.0)} \eqn\EqAmpsWithCrosscap $$
We also find
$F_{0}^{\prime (1/2)} ( \void ; A )
 = { 3^{1/2} \Gamma(1/4) \over 8 \pi^{3/2} } A^{-9/4}$
and
$F_{0}^{\prime (1)} ( \void ; A )
 = { 1 \over 3 \pi^{1/2} } A^{-1}$
from {\EqAmpsWithCrosscap}
while
$F_{0}^{(0)} ( \void ; A )
 = { 3 \over 16 \pi } A^{-7/2}$
and
$F_{0}^{(1)} ( \void ; A )
 = { 1 \over 36 \pi^{1/2} } A^{-1}$.
\par
In the rest of this paper, we investigate the non-critical string field theory
with matter fields on the world-sheet in the following two methods:
First, we consider the non-critical string field theory
which corresponds to
$m$-th multicritical one matrix model
($m \ge 2$).\NPrefmark{\Kazakov}
For $m \ge 3$, matter fields are incorporated
because the central charge is $c = - 2 (m-2) (6m-7) / (2m-1) < 0$.
The $m=2$ model corresponds to $c=0$ non-critical string theory.
In these models the lattice surfaces have some kinds of regular polygons
instead of only regular triangles.
After taking the continuum limit, we obtain the Hamiltonian,
$$ \eqalign{ \sp(2.0)
&\H (\cc_2,\ldots,\cc_m)
\cr\sp(4.0)
&= \  - \intdzeta \, \bigl\{ \,
 \omega (\zeta,\cc_2,\ldots,\cc_m) \, \Psi (- \zeta)
 \, + \, ( \Phi^\dagger (\zeta,\cc_2,\ldots,\cc_{m-1}) )^2
      \, {\der \over \der \zeta} \, \Psi (- \zeta)
\cr\sp(0.1)
&\phantom{= \  - \intdzeta \, \bigl\{ \, }
 + \, ( G + \Gxx ) \, \Phi^\dagger (\zeta,\cc_2,\ldots,\cc_{m-1})
                   \, ( {\der \over \der \zeta} \, \Psi (- \zeta) )^2
\cr\sp(0.1)
&\phantom{= \  - \intdzeta \, \bigl\{ \, }
 + \, \Gx \, \Phi^\dagger (\zeta,\cc_2,\ldots,\cc_{m-1})
          \, {\der^2 \over \der \zeta^2} \, \Psi (- \zeta) \, \bigr\} \, ,
\cr\sp(3.0)} \eqn\EqHKCon $$
where
$\omega(\zeta,\cc_2,\ldots,\cc_m) =
 {\der \over \der \zeta} ( F_1^{{\rm univ}(0)}(\zeta;\cc_2,\ldots,\cc_m) )^2$
and $F_1^{{\rm univ}(0)}(\zeta;\cc_2,\ldots,\cc_m)$
is the universal part of disk amplitude.
In the case of $m=2$, {\EqHKCon} is identical to {\EqHCon}.
The Hamiltonian {\EqHKCon} satisfies
the consistency condition which was discussed in ref.\ [{\IK}]
for any $m$ ($\ge 2$).
Second, let us consider to introduce matter fields
by putting a matter field $\phi$ on each link of the lattice surface.
Since matter fields are put on each link,
the wave function depends on
not only the length of string $l$
but also the values of matter fields on string,
i.e.,
$\Psi^\dagger = \Psi^\dagger ( l ; \phi_1 , \ldots , \phi_l )$.
The $\k_n$ for a regular $n$-polygon
also depends on matter fields on the link as
$\k_n = \k_n (\phi_1,\ldots,\phi_n)$.
{}From now on, we consider triangulated surfaces, for simplicity.
The extension to the case with $n$-polygons is straightforward.
If the cyclic conditions,
${\hat \Psi}^\dagger ( l ; \phi_1, \phi_2, \ldots, \phi_l )
 =
 {\hat \Psi}^\dagger ( l ; \phi_2, \ldots, \phi_l, \phi_1 )$
and
$\k ( \phi_1, \phi_2, \phi_3 )
 =
 \k ( \phi_2, \phi_3, \phi_1 )$,
are satisfied,
we obtain the Hamiltonian at the discrete level as follows:
$$ \eqalign{ \sp(2.0)
\Htot \ = \
& \sum_{l=1}^\infty \int \! d \phi_1 \cdots d \phi_l \, \bigl\{ \,
  {\hat \Psi}^\dagger ( l ; \phi_1, \ldots, \phi_l )
\cr\sp(4.0)
& - \int \! d \phi' d \phi'' \,
  \k (\phi_1,\phi',\phi'') \,
  {\hat \Psi}^\dagger ( l+1 ; \phi_2, \ldots, \phi_l, \phi', \phi'' ) \,
  \bigr\} \, l \, {\hat \Psi} ( l ; \phi_1 , \ldots , \phi_l )
\cr\sp(4.0)
& - \, \sum_{l=2}^\infty \int \! d \phi_1 \cdots d \phi_l \,
  \sum_{l'=0}^{l-2}
  \bigl\{ \,
    {\hat \Psi}^\dagger ( l' ; \phi_2, \ldots, \phi_{l'+1} ) \,
    {\hat \Psi}^\dagger ( l-l'-2 ; \phi_{l'+3}, \ldots, \phi_{l} )
\cr\sp(0.1)
& \phantom{- \, \sum_{l=2}^\infty \int \! d \phi_1 \cdots d \phi_l \,
           \sum_{l'=0}^{l-2} \bigl\{ \, }
    + \, \gx \,
    {\hat \Psi}^\dagger ( l-2 ; \phi_2, \ldots, \phi_{l'+1},
                         \phi_{l}, \ldots, \phi_{l'+3} )
  \, \bigr\}
\cr\sp(4.0)
& \phantom{- \, } \
  \times \,
  l \, {\hat \Psi} ( l ; \phi_1 , \ldots , \phi_l ) \,
  \delta ( \phi_1 - \phi_{l'+2} )
\cr\sp(6.0)
& - \, \sum_{l=1}^\infty \sum_{l'=1}^\infty
  \int \! d \phi_1 \cdots d \phi_l \,
          d \tilde \phi_1 \cdots d \tilde \phi_{l'} \,
  \bigl\{ \,
    g \, {\hat \Psi}^\dagger ( l+l'-2 ; \phi_2, \ldots, \phi_l,
                                 \tilde \phi_2, \ldots, \tilde \phi_{l'} )
\cr\sp(0.1)
& \phantom{- \, \sum_{l=1}^\infty \sum_{l'=1}^\infty
           \int \! d \phi_1 \cdots d \phi_l \,
                   d \tilde \phi_1 \cdots d \tilde \phi_{l'} \, \bigl\{ \,
           \hskip -19pt }
    + \,
    \gxx \, {\hat \Psi}^\dagger ( l+l'-2 ; \phi_2, \ldots, \phi_l,
                                    \tilde \phi_{l'}, \ldots, \tilde \phi_2 )
  \, \bigr\}
\cr\sp(4.0)
& \phantom{- \, } \
  \times \,
  l \,  {\hat \Psi} ( l  ; \phi_1, \ldots, \phi_l ) \,
  l' \, {\hat \Psi} ( l' ; \tilde \phi_1, \ldots, \tilde \phi_{l'} ) \,
  \delta ( \phi_1 - \tilde \phi_1 ) \, ,
\cr\sp(3.0)} \eqn\EqHMatter $$
where the integration $\int\! d \phi$
is replaced by the summation $\sum_\phi$
if the matter fields take discrete values.
Since one can introduce any kinds of matter fields in {\EqHMatter},
one obtains the non-critical string field theory with any value of $c$
(including $c > 1$ case) at the discrete level.
However, we have not succeeded in
taking the continuum limit of {\EqHMatter} yet.
%
%
\par
In conclusion,
in this paper we have investigated the non-critical string field theory
with non-orientable string interactions,
according to the systematic method by ref.\ [{\Biki}].
We have introduced two coupling constants $G$ and $\Gx$
which satisfy the relation $(\Gx)^3 = G \Gx$,
where $G$ counts the number of untwisted handles of the world-sheet,
while $\Gx$ counts the number of cross-caps.
In the beginning
we have constructed the $c=0$ non-critical string field theory.
By using the transfer matrix formalism in the dynamical triangulation,
we have formulated the $c=0$ string field theory at the discrete level,
and then we have obtained the Hamiltonian at the continuous level
{\EqHCon} with {\EqOmegaCon}
after taking the continuum limit.
The continuum limit is taken by {\EqConLimitxyk} $\sim$ {\EqConLimitG}
so as to satisfy the condition $\dim{\T} > 0$.
Any transfer matrices and any amplitudes are calculated by
{\EqDefPhiDaggerCon} $\sim$ {\EqGenFunCon} and {\EqGenAmpCon}.
For example, we have calculated
some amplitudes with cross-caps on the world-sheet.
We have also studied the universality by showing that
some modified Hamiltonians at the discrete level always lead to
the same Hamiltonian {\EqHCon} with {\EqOmegaCon}
after taking the continuum limit.
Moreover,
we have extended our formalism to the string field theory with matter fields.
As one of extensions we have studied the non-critical string field theory
which corresponds to $m$-th multicritical one matrix model ($m = 2,3,\ldots$).
We have succeeded in taking the continuum limit
and have found the Hamiltonian {\EqHKCon}.
As another example,
we have put a matter field naively on each link of the lattice surface.
At the discrete level, we have obtained the Hamiltonian {\EqHMatter}
with any value of $c$ (including $c > 1$ case).
Taking the continuum limit of {\EqHMatter} is the future problem.
\NextPage
%
%
%
\ack
The author would like to thank
Dr.\ N.\ Kawamoto and Dr.\ S.\ Nishigaki
for useful discussions.
The work of the author is supported in part by Soryushi Shogakukai.
\NextPage
%
%
\def\Figures{
\centerline{\BIGr FIGURE CAPTIONS}
\vskip 8pt
\item{\rm Fig.~1}{
  Decomposition of a surface by peeling
  }
\item{\rm Fig.~2}{
  Five basic minimal-step \lq peeling decompositions'
  where a solid line and a broken line represent
  an initial string and a final string, respectively.
  In Fig.~2a a triangle is removed
  while in the rest of Figs.\ a two-folded part is removed.
  }
\item{\rm Fig.~3}{
  Ten basic minimal-step \lq peeling decompositions'
  without introducing the two-folded parts.
  A solid line and a broken line represent
  an initial string and a final string, respectively.
  In all Figures one triangle is removed.
  Two of links of each triangle in Fig.~3g and Fig.~3h
  are glued together so as to make no cross-cap and one cross-cap,
  respectively.
  }
          }
%
%
\refout
\NextPage
\Figures
\vfill
%
\bye